# A parameter-free method to extract the superconductor's $J_c(B,\theta)$ field-dependence from in-field current-voltage characteristics of HTS tapes

Víctor M. R. Zermeño[1], Krzysztof Habelok[2], M. Stępień[2] and Francesco Grilli[1]

[1]Institute for Technical Physics Karlsruhe Institute of Technology, Karlsruhe, Germany

[2]Department of Power Electronics, Electrical Drives and Robotics Silesian University of Technology, Gliwice, Poland

## Abstract

The estimation of the critical current ($I_c$) and AC losses of high-temperature superconductor (HTS) devices through modeling and simulation requires the knowledge of the critical current density ($J_c$) of the superconducting material. This $J_c$ is in general not constant and depends both on the magnitude ($B_{loc}$) and the direction ($\theta$, relative to the tape) of the local magnetic flux density. In principle, $J_c(B_{loc}\,\theta)$ can be obtained from the experimentally measured critical current $I_c(B_a,\theta)$, where $B_a$ is the magnitude of the applied magnetic field. However, for applications where the superconducting materials experience a local field that is close to the self-field of an isolated conductor, obtaining $J_c(B_{loc},\theta)$ form $I_c(B_a,\theta)$ is not a trivial task. It is necessary to solve an inverse problem to correct for the contribution derived from the self-field. The methods presented in the literature comprise a series of approaches dealing with different degrees of mathematical regularization, such as the use of brute force or optimization methods to fit the parameters of preconceived non-linear formulas. In this contribution, we present a parameter-free method that provides excellent reproduction of experimental data and requires no human interaction or preconception of the $J_c$ dependence with respect to the magnetic field. In particular, it allows going from the experimental data to a ready-to-run $J_c(B_{loc},\theta)$ model in a few minutes.

## 1. Introduction

Most numerical models used to calculate the critical current or the AC losses of superconducting devices need an expression for $J_c(B_{loc},\theta)$ as constitutive law of the superconductor material. Here $B_{loc}$ is the local magnetic flux density experienced by the superconductor. The most obvious way of obtaining such $J_c(B_{loc},\theta)$ dependence is to take the $I_c(B_a, \theta)$ values measured on short tape samples (where $B_a$ is the magnitude of the *applied* field) and divide them by the superconductor's cross-section. The obtained data can then be fitted with an analytical formula (or simply interpolated) and successively used for simulating devices made of that tape. This approach may work, but not for low applied fields: in fact, the experimentally measured $I_c(B_a,\theta)$ includes the contribution of the self-field generated by the transport current during the measurement of the current-voltage characteristics. If the applied field is not very large, the self-field contribution can be comparable to the applied field. In other words $B_{loc}$ is not equal to $B_a$. This can lead to inaccuracies in successive calculations. The influence of the self-field needs therefore to be properly "subtracted" from the experimental data.

Several methods have been proposed to solve this problem: Rostila *et al* [1], Pardo *et al* [2], Zhang *et al* [3], Gomory *et al* [4], and Grilli *et al* [5] have managed to find expressions that allow reproducing the observed angular dependencies. However, they all make use of preconceived analytic formulas for the $J_c(B_{loc},\theta)$ relation. Finding such formulas can be very time consuming, especially for tapes with artificial pinning centers, which exhibit a rather complex angular dependence. In some cases 10 or more parameters may be necessary for a sufficiently accurate reproduction of the experimental data [2,6]. This complexity has been observed also at low temperatures and high fields [7,8].

In this contribution, we present a new approach that allows extracting a local $J_c(B_{loc},\theta)$ model from experimental $I_c(B_a,\theta)$ data without the need of using preconceived analytical formulas nor fitting parameters.

The method solves the forward problem by considering a numerical approximation to $J_c(B_{loc},\theta)$ in the form if an interpolation function. The computed critical current values are compared to the experimental $I_c(B_a,\theta)$ data and the error is fed back to modify the interpolation function used for $J_c(B_{loc},\theta)$. The process is repeated iteratively, creating every time a new interpolation function for $J_c(B_{loc},\theta)$, until the difference between the computed critical current values and the experimental $I_c(B_a,\theta)$ data reaches a minimum.

The examples discussed in this paper refer to HTS coated conductors, but the methods discussed here are valid for other HTS tapes as well, and in general for any superconductor operating in conditions where the magnitude of the self-field is comparable to that of the external applied field.

## 2. Method description

This section first describes the common approaches used to extract the $J_c(B_{loc}, \theta)$ dependence from $I_c(B_a, \theta)$ values measured on short tape samples. Then, after emphasizing the limitations of those approaches, it describes a new proposed method, which does not require the knowledge of any preconceived formula for the angular dependence.

Once the $J_c(B_{loc}, \theta)$ is extracted (with any method), one can test its correctness by calculating the critical current of the tape in self-field and in external field (of different amplitudes and orientations) and compare the results to the experimental values. We call this the solution of a *forward* problem, in opposition to the *inverse* problem of determining the local critical current density $J_c(B_{loc}, \theta)$ from the tape's critical currents $I_c(B_a, \theta)$.

In this paper, in order to solve the forward problem, we solve the following equation (as proposed in [9,10])

$$\nabla \times (\nabla \times A) = J_c(B_{loc}, \theta) \qquad (1)$$

Solving (1) means finding the self-consistent magnetic flux density distribution such that all the points of the superconductor are at their critical current density $J_c(B_{loc}, \theta)$, which is of course different in every point.

### 2.1. Common techniques used to extract the angular dependence

The simplest approach to obtain the $J_c(B_{loc}, \theta)$ dependence consists in taking the experimental data and dividing them by the cross-section of the superconductor. This leads to inaccuracies at low fields and in particular to an underestimation of the critical current when the latter is recalculated with the forward model.

Figure 1 shows an example for a 4 mm coated conductor with a self-field critical current of about 160 A. In this and the similar figures in this paper, the angle $\theta$ is defined with respect to the c-axis of the superconductor: so 0° and 90° mean a magnetic field perpendicular and parallel to the flat face of the tape, respectively. The average self-field at $I_c$=160 A is about 20 mT. As a consequence, the error caused by this approach is already small for a background field of 100 mT and vanishes for fields of 200 mT. However, in applications like HTS cables for power transmission and bifilar coils for resistive fault current limiters the superconductor experiences field of only a few tens of mT, so the inaccuracy introduced by this method plays an important role.

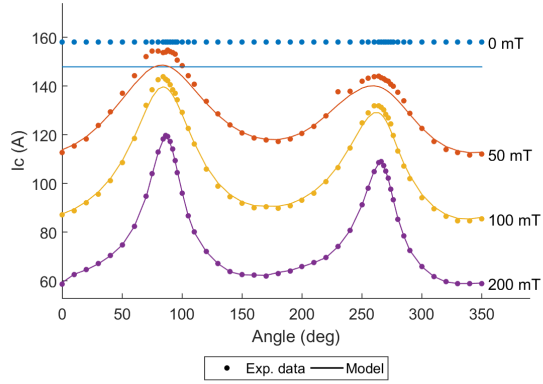

**Figure 1.** Continuous lines: $I_c(B_a, \theta)$ angular dependence calculated by using $J_c(B_{loc}\ \theta)=I_c(B_a,\theta)/S$ as input for the forward problem, where S is the cross-section of the superconductor. Solid points: experimental data. The calculation gives inaccurate results in self-field and at low fields (50 mT).

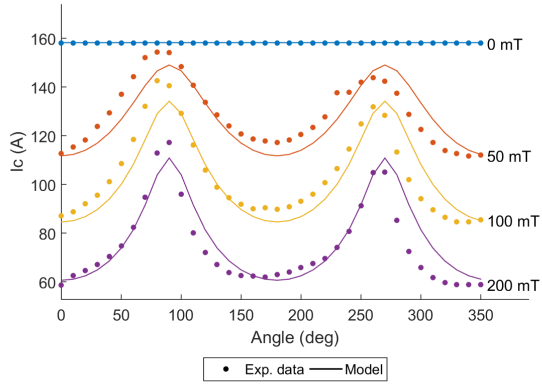

**Figure 2.** Continuous lines: $I_c(B_a, \theta)$ angular dependence calculated by using an analytic formula for $J_c(B_{loc}\ \theta)$ as input for the forward problem, with the best set of parameters. Solid points: experimental data. The calculation gives the correct value in self-field, but cannot exactly reproduce the angular dependence in applied field. Given the lack of symmetry of experimental data, more parameters would be needed in the expression for $J_c(B_{loc}\ \theta)$.

An alternative approach, which can take the self-field effects into account, consists in determining the parameters of a preconceived formula for the angular dependence of $J_c$. The formula is chosen based on the appearance of the measured angular dependence of $I_c$. For example, in the case considered here, the critical current exhibits an elliptical dependence, with two peaks corresponding to a field applied parallel to face of the tapes. The elliptical dependence is able to describe the angular dependence of $J_c$ caused by flux pinning in polycrystalline high-temperature superconductors in low magnetic fields (up to a few hundreds of mT) [11]. It is typically described by a four-parameter formula:

$$J_c = \frac{J_{c0}}{\left(1 + \frac{\sqrt{(kB_\parallel)^2 + {B_\perp}^2}}{B_c}\right)^b} \qquad (2)$$

where $B_\parallel$ and $B_\perp$ are the magnetic field components parallel and perpendicular to the flat face of the tape, respectively, $k$ is the parameter describing the degree of anisotropy, $B_c$ is a magnetic field value to describe how quickly $J_c$ decreases, and $b$ is a smoothing factor.

A formula such as (2), however, cannot take into account the fact that the measured data of Figure 1 do not have a 180° symmetry. For a more accurate description of the angular dependence, formulas with more parameters need to be used. For the purpose of illustrating the method, however, we consider here an elliptic dependence.

In order to determine the best parameters for the $J_c(B_{loc},\theta)$ dependence, a "brute-force" approach has been proposed [5]. The approach consists in choosing a range of variation for the different parameters, and solving the forward problem for all the possible combinations. The calculated angular dependence of the critical current is then compared to experimental data, and the "best" set of parameters that minimizes the error with respect to the experimental data is then chosen.

The brute-force approach requires solving a large number of forward problems, one for each combination of parameters, field amplitude and orientation. In [5], the authors started from a symmetrized version of the experimental data (so that only angles between 0 and 90° need to be considered) and solved 52,920 forward problems to determine the best parameters. The solution of an individual forward problem by means of 2-D finite-element simulations is quite fast, about one second, but this still results in several hours of computation. In the specific case of coated conductors, considering the superconductor as a 1-D object can dramatically shorten the simulation time. The same 52,920 cases presented in [5] could be solved in 130 s, with the 1-D forward problem implemented in Matlab.

Although this brute force approach allows obtaining a relatively small error with respect to the experimental data (relative error of 1.95% averaged on all measured data points, see Figure 2), one is not sure whether the "best" set of parameters found in this way is really the one minimizing the error. The absolute minimum of the error most probably corresponds to values of the parameters different from those considered in the discrete range of variation. A more efficient search for the set of parameters that minimizes the error with respect to experimental data can be performed by means of dedicated algorithms, such as the Nelder-Mead [12] and the MMA algorithms [13].

The results of the various approaches are detailed in [14] and summarized in Table 1. As reported in [5], relatively different values of the parameters can provide a similarly good reproduction of the experimental data.

Table 1. Comparison of results and computation times obtained with different methods using an analytic formula for the $J_c(B_{loc},\theta)$ dependence shown in equation (2).

|  | $J_{c0}$ (A m$^{-2}$) | $B_c$ (T) | $b$ | $k$ | Error (%) | Time (s)* |
|---|---|---|---|---|---|---|
| **Comsol (2-D)** | | | | | | |
| Brute-force | 4.75 × 10$^{10}$ | 0.035 | 0.6 | 0.25 | 1.95 | 36000 |
| Nelder-Mead | 4.95 × 10$^{10}$ | 0.031 | 0.59 | 0.277 | 1.64 | 2400 |
| MMA | 4.63 × 10$^{10}$ | 0.055 | 0.74 | 0.287 | 1.88 | 3600 |
| **Matlab (1-D)** | | | | | | |
| Brute-force | 4.75 × 10$^{10}$ | 0.035 | 0.6 | 0.25 | 1.95 | 130 |
| Nelder-Mead | 4.56 × 10$^{10}$ | 0.056 | 0.73 | 0.274 | 1.59 | 20 |

* Times refer to a desktop workstation with Inter Core i7 processors at 3.3 GHz and 64 GB of RAM memory.

All the methods considered so far are based on the use of a preconceived formula for the angular dependence of $J_c$. In the case of complex angular dependencies, such as those presented in [2,15], the mere choice of the analytical formula to use is not trivial and requires a lot of manual tweaks (see for example the appendix of [2]). In addition, samples labeled as "similar" by the manufacturers often exhibit different angular dependencies. In those cases, the procedure for finding the analytic formula has to be started from scratch every time, which makes this approach very time consuming.

### 2.2. A parameter-free method

In order to overcome the shortcomings of the methods described in the previous section, we developed a parameter-free method, whose output is a series of data points describing the $J_c(B_{loc},\theta)$ dependence. These data points can be used as input for successive simulations, e.g. for calculating the critical current or AC losses of superconducting devices. The model used for these successive simulations must be able to import the data and interpolate them.

The steps of the procedure for extracting the $J_c(B_{loc},\theta)$ dependence are listed here below and schematically illustrated in the flow diagram of Figure 3:

- The initial $J_c(B_{loc},\theta)$ data are simply the experimental $I_c(B_a,\theta)$ values divided by the cross section of the superconductor.
- The forward model is then run for all the magnitudes and orientations of the applied field. This produces a first set of calculated critical currents (one for each value of field amplitude and orientation), which are then compared against the experimentally measured ones.

- The error between calculated and measured critical current is evaluated and then smoothed to rule out for outliers, which could be due to experimental error and to avoid overfitting.
- The smoothed error is divided by the cross section of the superconductor and subtracted to the present $J_c(B_{loc},\theta)$ interpolating function, hence providing a new $J_c(B_{loc},\theta)$ estimate.
- The procedure is repeated until the error between the calculated critical currents and the experimentally measured ones is sufficiently small.

A key step in this procedure is the smoothing of the error, which is illustrated in Figure 4. The figure shows how the $J_c(B_{loc},\theta)$ changes from the first iteration of the process (Figure 4a) to the second one (Figure 4c) and how the error is smoothed in between (Figure 4b). In Figure 4a the initial distribution presents a plateau at small fields (around $B$=0, i.e. near the center), because this region is dominated by the self-field. In Figure 4b the lower field curve (50 mT) is the one with the larger error with respect to the experimental data. The error is then smoothed by means of Matlab's `smooth` function, with the option `rloess`[1] and a span parameter of 0.3. The span parameter determines the fraction of neighboring elements in the data set that contribute to the smoothing of every point.

Already after just one iteration the distribution of $J_c(B_{loc},\theta)$ became more peaked, as it can be seen from the darker area that appears in the center of Figure 4c: this is because the contribution of the self-field has been subtracted. No significant changes occur in the regions far from the center, corresponding to larger field amplitudes, where the self-field is negligible.

---

[1] See Matlab's documentation: http://www.mathworks.com/help/curvefit/smooth.html

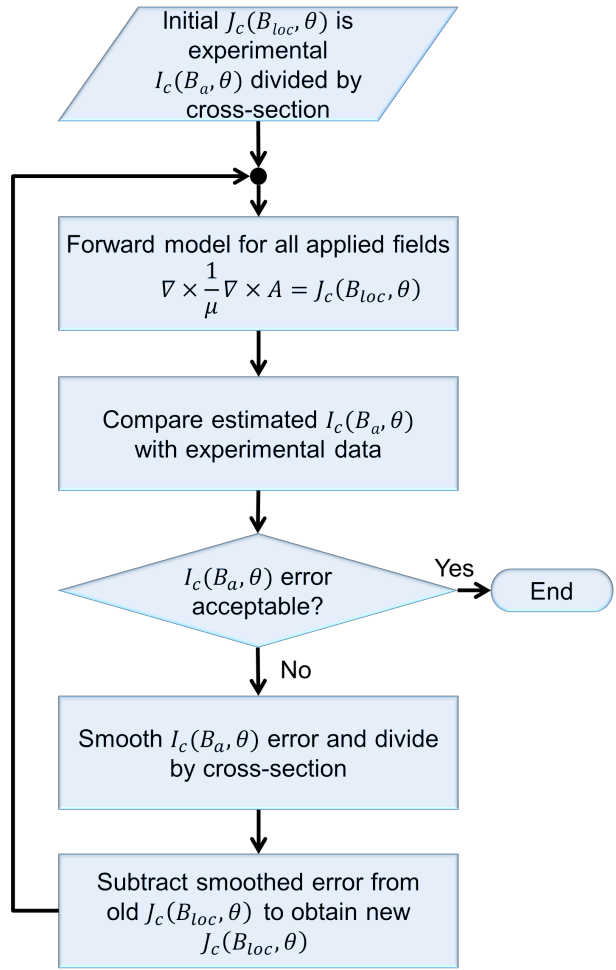

**Figure 3. Flow chart of the parameter-free procedure.**

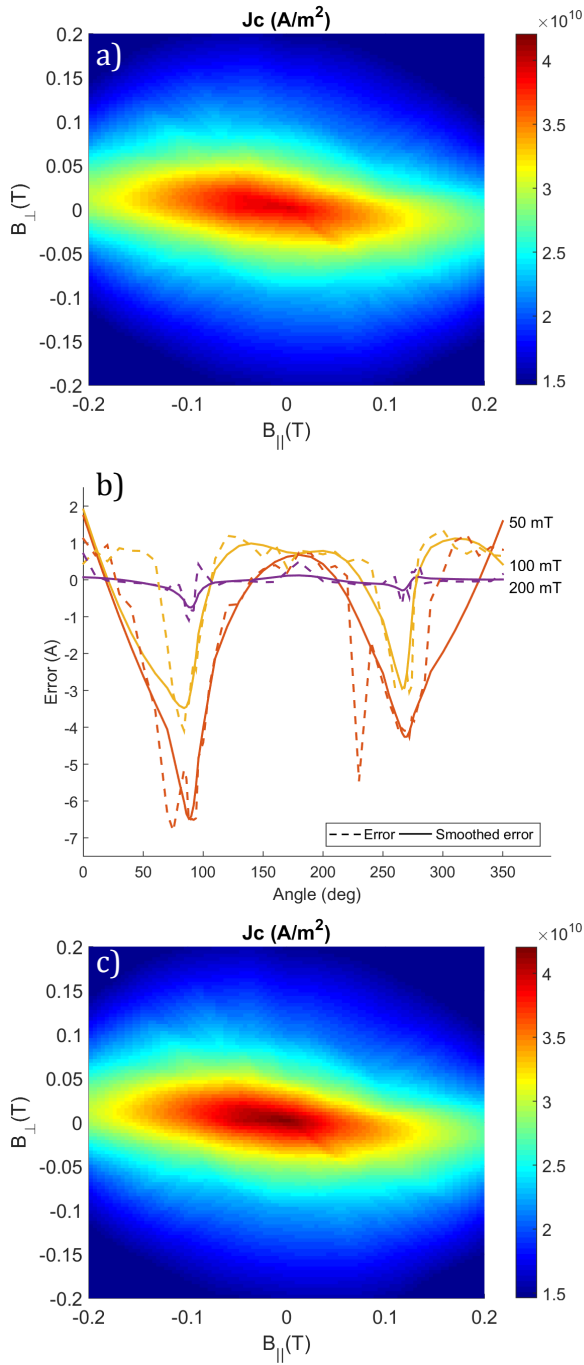

**Figure 4.** Details of the first two iteration of the parameter-free method. (a) In the initial iteration $J_c(B_{loc},\theta)=I_c(B_a,\theta)/S$ is used for the forward problem (S is the superconductor's cross-section and $I_c(B_a,\theta)$ here is the measured angular dependence), (b) The angular dependence calculated solving the forward problem is compared to the experimental data. The error is evaluated and smoothed. (c) A new $J_c(B_{loc},\theta)$ is obtained. It will be inserted in the forward problem and the procedure will be repeated until the error goes below a pre-determined value.

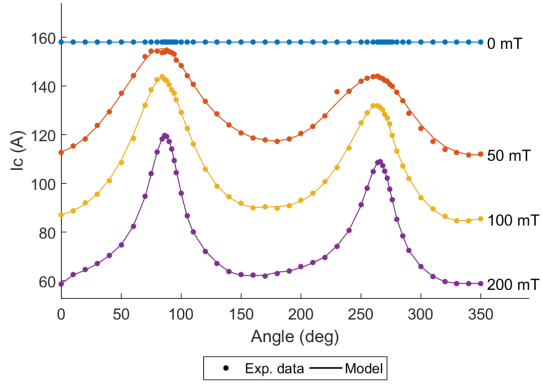

**Figure 5.** Continuous lines: $I_c(B_a, \theta)$ angular dependence calculated with the parameter-free method. Solid points: experimental data. The calculation gives the correct value for all field values (including self-field) and is able to precisely reproduce the angular dependence. Note that a couple of outlying points of the 50 mT curve are discarded.

The advantage of this approach is that no formulas for $J_c(B_{loc},\theta)$ are needed. In other words, one manipulates only sets of data points: the experimental $I_c(B_a,\theta)$ are divided by the cross section and properly modified to take the self-field effects into account. The final modified set of data, divided by the cross section, constitutes the local $J_c(B_{loc},\theta)$ dependence that one can use as a superconductor's property in successive simulations.

This procedure has been implemented both for a 2-D and 1-D description of the tape in Comsol Multiphysics and Matlab, respectively. The obtained $J_c(B_{loc},\theta)$ data set are slightly different, due to the different physical description of 2-D and 1-D conductors (for example, no variation of the electromagnetic quantities along the tape thickness is possible in 1-D). As a consequence, the $J_c(B_{loc},\theta)$ data set cannot be interchanged between the 2-D and 1-D models .

The agreement between the $I_c(B_a, \theta)$ angular dependence calculated with this method and experimental data is excellent, as shown in Figure 5: the model nicely reproduces the experimental values for all field amplitudes and orientations (including self-field) and discards the obvious outlying points present in the 50 mT data set.

## 3. Application of the method and discussion

The parameter-free approach has been applied to tapes exhibiting higher critical current and/or more complex angular dependence. Three examples are reported in Figure 6, which shows the angular $I_c(B_a, \theta)$ dependence for a 4 mm wide tape with artificial pinning centers, and two 12 mm wide samples without and with artificial pinning centers, respectively. In the figure, the solid points represent the measured values. The continuous lines represent the critical currents calculated by means of the parameter-free method. In all the considered examples, the parameter-free method is able to reproduce the experimental data with a mean error (averaged on all data points) of less than 0.15%. The agreement with experimental data is much better than what is typically obtained with analytical formula, as can be seen by comparing these results with those of Figure 2 and 3 of [6].

A summary of the advantages and disadvantages of the different methods to extract the $J_c(B_{loc},\theta)$ dependence is presented in Table 2. As shown by the results of this paper, the parameter-free method has clear advantages compared to the direct use of experimental data and analytic formulas – both in terms of accuracy and speed of implementation and use of the method. The only drawback of the parameter-free method is that, especially in case of rapidly varying $J_c(B_{loc},\theta)$ dependencies, it can lead to the model overlooking said variations if a strong smoothing is applied. On the other hand, lack of smoothing will lead to over-fitting. This potential problem could however be solved by appropriately changing the span parameter in the smoothing function. In all the examples presented in this work the same span parameter (0.3) was used without problem.

**Table 2. Advantages and disadvantages of the different methods for extracting $J_c(B_{loc},\theta)$**

|  | Direct use of experimental data | Analytic formula | Parameter-free method |
|---|---|---|---|
| **Advantages** | No processing required | Good for general examples | No user input needed |
|  | Acceptable for large–field applications | Can give insight into material's properties | Computationally very fast when compared to parameter fitting in analytic formulas |
|  | No regularization |  | Low regularization |
| **Disadvantages** | Not good for low-field applications | Requires development of analytical formula | If error is not smoothed properly, it can lead to over-fitting |
|  |  | Parameter estimations leads in general to long computation times |  |
|  |  | High regularization |  |

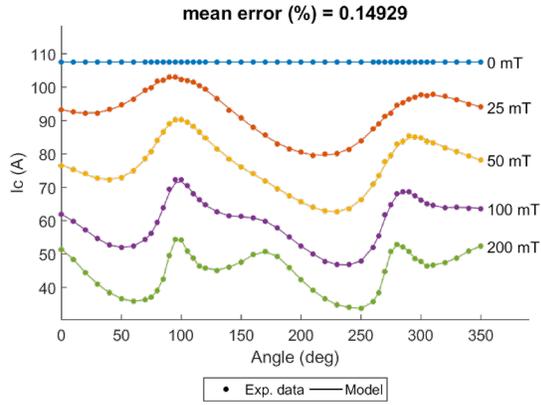

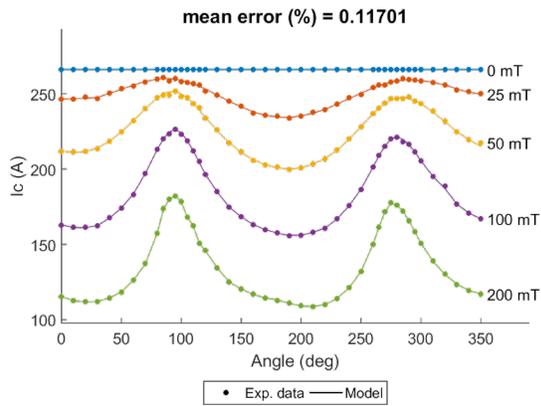

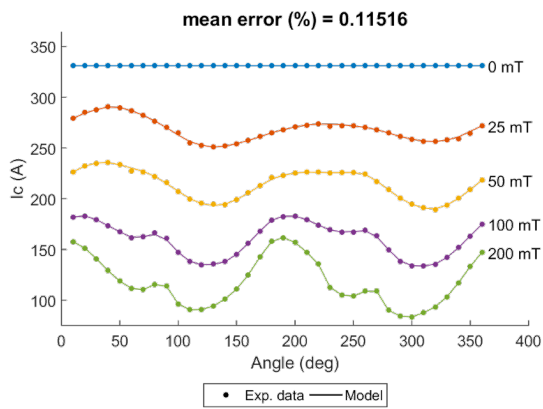

Figure 6. Measured (data points) and calculated (continuous lines) $I_c(B_a, \theta)$ dependence for different tape samples: (a) 4 mm wide tape with artificial pinning centers (SuperPower); b) 12 mm wide tape without artificial pinning centers (SuperOx); c) 12 mm wide tape with artificial pinning centers (SuperPower).

## 4. Conclusion

In this contribution, a parameter-free method to extract the local angular dependence of the critical current density of the superconductor material from in-field voltage-current characteristics of short HTS tape samples is proposed. The method properly accounts for the self-field produced by the sample during experimental measurements of the critical current. Differently from other methods proposed in the literature, it does not make use of analytic formulas for the description of the angular dependence of $J_c$ – a particularly welcome feature with tapes with artificial pinning centers exhibiting complex angular dependencies. The output of the method is a set of $J_c(B,\theta)$ data that, once interpolated, can be used as material's properties for successive simulations of HTS devices. The proposed method is very fast, and typically allows going from the experimental data points to the $J_c(B,\theta)$ in a matter of minutes with no human interaction needed.

## References

[1] Rostila L, Lehtonen J, Mikkonen R, Šouc J, Seiler E, Melíšek T and Vojenčiak M 2007 How to determine critical current density in YBCO tapes from voltage–current measurements at low magnetic fields *Supercond. Sci. Technol.* **20** 1097–100

[2] Pardo E, Vojenčiak M, Gömöry F and Šouc J 2011 Low-magnetic-field dependence and anisotropy of the critical current density in coated conductors *Supercond. Sci. Technol.* **24** 65007

[3] Zhang M, Kvitkovic J, Pamidi S V and Coombs T A 2012 Experimental and numerical study of a YBCO pancake coil with a magnetic substrate *Supercond. Sci. Technol.* **25** 125020

[4] Gomory F, Souc J, Pardo E, Seiler E, Soloviov M, Frolek L, Skarba M, Konopka P, Pekarcikova M and Janovec J 2013 AC Loss in Pancake Coil Made From 12 mm Wide REBCO Tape *IEEE Trans. Appl. Supercond.* **23** 5900406

[5] Grilli F, Sirois F, Zermeno V M R and Vojenciak M 2014 Self-Consistent Modeling of the Ic of HTS Devices: How Accurate do Models Really Need to Be? *IEEE Trans. Appl. Supercond.* **24** 8000508

[6] Grilli F, Vojenciak M, Kario A and Zermeno V 2016 HTS Roebel Cables: Self-Field Critical Current and AC Losses Under Simultaneous Application of Transport Current and Magnetic Field *IEEE Trans. Appl. Supercond.* **26** 8000508

[7] Long N J 2008 Model for the angular dependence of critical currents in technical superconductors *Supercond. Sci. Technol.* **21** 025007

[8] Hilton D K, Gavrilin A V and Trociewitz U P 2015 Practical fit functions for transport critical current versus field magnitude and angle data from (RE)BCO coated conductors at fixed low temperatures and in high magnetic fields *Supercond. Sci. Technol.* **28** 074002

[9] Majoros M, Glowacki B A and Campbell A M 2001 Critical current anisotropy in Ag/(Pb,Bi)$_2$Sr$_2$Ca$_2$Cu$_3$O$_{10+x}$ multifilamentary tapes: influence of self-magnetic field *Supercond. Sci. Technol.* **14** 353–62

[10] Gömöry F and Klinčok B 2006 Self-field critical current of a conductor with an elliptical cross-section *Supercond. Sci. Technol.* **19** 732–7

[11]   Gömöry F, Šouc J, Vojenčiak M and Klinčok B 2007 Phenomenological description of flux pinning in non-uniform high-temperature superconductors in magnetic fields lower than the self-field *Supercond. Sci. Technol.* **20** S271

[12]   Nelder J A and Mead R 1965 A Simplex Method for Function Minimization *Comput. J.* **7** 308–13

[13]   Svanberg K 1987 The method of moving asymptotes—a new method for structural optimization *Int. J. Numer. Methods Eng.* **24** 359–73

[14]   Habelok K 2015 *Development of numerical models for extracting the Jc(B) angular dependence of HTS wires from experimental data* (Silesian University of Technology, Gliwice, Poland) Available at https://bwsyncandshare.kit.edu/dl/fiTp9X4R884PMQvwQ6AcB2QX/Habelok_Master_Thesis.pdf

[15]   Vojenčiak M, Grilli F, Terzieva S, Goldacker W, Kováčová M and Kling A 2011 Effect of self-field on the current distribution in Roebel-assembled coated conductor cables *Supercond. Sci. Technol.* **24** 095002